# Exploring the socio-technical interplay of Industry 4.0: a single case study of an Italian manufacturing organisation

Emanuele Gabriel **Margherita**[a], Alessio Maria **Braccini**[a]

[a] *University of Tuscia, Department of Economics Engineering Society and Organization – DEIM, Via del Paradiso, 47, 01100, Viterbo, VT, Italy University*

**Abstract**

In this position paper, we explore the socio-technical interplay of Industry 4.0. Industry 4.0 is an industrial plan that aims at automating the production process by the adoption of advanced leading-edge technologies down the assembly line. Most of the studies employ a technical perspective that is focused on studying how to integrate various technologies and the resulting benefits for organisations. In contrast, few studies use a socio-technical perspective of Industry 4.0. We close this gap employs the socio-technical lens on an in-depth single case study of a manufacturing organisation that effectively adopted Industry 4.0 technologies. The findings of our studies shed light both on the socio-technical interplay between workers and technologies and the novel role of workers. We conclude proposing a socio-technical framework for an Industry 4.0 context.

## 1. Purpose of Research

Industry 4.0 (I40) is an industrial initiative that aims at improving the production process through the adoption of leading-edge technologies such as robotics, the Internet of Things, big data analytics and cloud manufacturing [1, 2]. I40 technologies improve the efficiency and flexibility of production processes and deliver economic value in the form of higher productivity, higher-quality products [3, 4], better logistics performance [5, 6], and enhanced inventory management [7]. The integration of these advanced technologies enables the deployment of the programmable interconnected cyber-physical systems, that drive fully automatic pieces of machinery used in the assembly lines allowing to address problems on the assembly line without human interaction but through autonomous machines that take on the job from the workforce [8]. Since I40 initiatives have as a primary focus on the application of these technologies [9], studies consider the technologies as they are the main driver to produce goods in an I40 context [10]. In contrast, the role of the workforce in an I40 production line is increasingly marginal as the I40 automation is not limited to production activities. Still, it is extended to decision-making activities which so far were prerogative of the workforce [11].

However, the application of the lens of socio-technical theory by Bostrom and Heinen [12] (Figure 1) – which considers the production process of an organisation as a work-system requiring to be conjointly optimised to operate - revealed that I40 also has implications on the social systems regarding the role of workforce and unit interacting within the production process.

Thus, our study aims at providing a socio-technical perspective that lacks in literature. Our study wants to investigate the socio-technical interplay with which I40 work systems afford to produce goods delivering value and the role of I40 workforce.

In doing so, we performed a single case study [13] of Italian Manufacturing organisations that have successfully adopted I40 applying the traditional socio-technical theory lens (Figure 1). The

---

study addresses the following research question: *"What is the socio-technical interplay in Industry 4.0 work systems?"*

We want to contribute to the literature of the socio-technical discourse applying the socio-technical theory on an I40 adoption, showing the socio-technical interplay of I40, and exploring the novel role of workers. We also present a socio-technical framework for I40, and we propose avenues to advance the socio-technical perspective of I40.

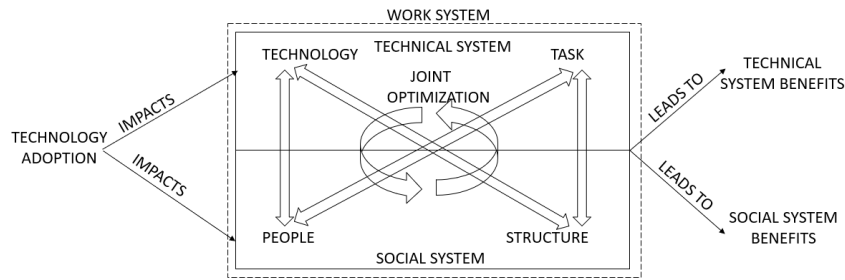

Figure 1 Socio-Technical Framework based on [12, 14]

## 2. Industry 4.0

I40 is an industrial initiative, started in Germany in 2011. It aims at ensuring the competitiveness of the manufacturing sector, which is undermined by manufacturers from developing countries selling products of good quality and at a lower price [14]. In doing so, a set of advanced technologies – Big Data, Internet of Things, Robotics, 3d printings – are adopted and integrated down the assembly line and allow the deployment of programmable, interconnected cyber-physical systems that control machinery automatically down the assembly lines. The way in which organisations use I40 technologies allows addressing problems on the assembly line without human interaction through autonomous machines [8]. Most of the literature of I40 is focused on the technical aspects of the initiative. Some studies are focused on the integration of these technologies [15–17]. Other studies focus on the economic benefits that I40 adoption delivers to manufacturing organisations. The literature revealed that I40 improved production operations making the production process more efficient and flexible [5, 14]. Natural resources are better used in an I40 production process because technologies perform more accurate operation [3].

Nonetheless, there is a lack of studies of I40 adoption with a socio-technical perspective. Few studies highlighted the impact of these technologies on workers in the form of a better work environment, reduction of strenuous tasks, increased autonomy for workers [18–20]. Further studies, using the term Industry 5.0, propose an I40 vision where workers can be co-creator with I40 technologies of value for organisations. However, they are in an infant stage, and more studies are required to validate this vision [9, 21]. Therefore, there is a lack of studies that explore how I40 technologies and workers interact and the novel role of workers in an I40 context because the initiative has a "technocentric" perspective. This perspective considers workers with a marginal role in the novel production process, and the automation of these technologies may degrade competencies, expertise and know-how of both blue and white-collar jobs [11, 14, 22].

## 3. Research Design

A single case is an appropriate research method to study a novel phenomenon that is not deeply investigated [13]. We chose the case for the investigation as the organisation is a pioneer in its sector to adopt I40 technologies. Also, the organisation provided us the opportunity to study the phenomenon deeply, and this is a requirement for a single case study [13]. To investigate the phenomenon, we had the chance to interview key informants [23], that is, the CEO and CPO who acted as the steering committee of the I40 adoption and two workers who operated in the previous tradition line and the novel I40 line. We conducted semi-structured interviews. The interview track was focused on the following themes: (I) the traditional production process and the role of the workers (II) the adoption of I40 technologies (III) the I40 production process and the role of workers. The interview track was adapted according to each interviewee.

We increase the reliability of the study triangulating secondary data in the form of web articles regarding the organisation and the adoption of I40 technologies [24]. We chose on reliable web articles which are retrieved from national newspapers. We analysed the data following the procedure for a qualitative inquiry by Corbin and Strauss [25]. The analysis was performed to identify the socio-technical interplay in I40 work systems.

## 4. Case Description

The case is a sanitary ceramic producer located in Italy with around 200 employees. The work systems embrace the traditional production process of the firm, which is labour-intensive. Workers manually operate machines and move products and materials through the assembly line. The mismanagement in materials handling introduced by the workforce errors reduces the quality of the product or increases the defect rate.

*"This is the traditional way of producing. Here the workers manually perform all the operation. They move the goods, crafts the bathroom and control the oven" (CPO).*

*"Before, the work was very hard. All the operations are muscular and very intense" (Worker 1).*

*"The traditional system of manufacture is labour intensive; workers operate with technologies" (Web article).*

Therefore, the management decided to adopt the following I40 technologies: self-driving forklifts and fully automatic conveyors to move the goods; automated robotised arms to produce the goods and Internet of Things application to integrate the technologies. Management discussed the choice to adopt these technologies with workers in order to improve their conditions and increase the quality of products. The worker's reaction was positive. Management and technology developers provide vocational training for workers to acquire proper skills and digital competencies to deal with technology. Moreover, management leaves in function the traditional production process for some productions, and it is used for apprentices to learn knowledge in learning ceramics.

*"We were the first in our sector to adopt these new technologies. After, several courses were provided to the workers because these technologies change a lot compared to the previous technologies" (CEO).*

*"We decided to adopt these technologies to improve our processes and make the tasks easier for our workers. These technologies can produce and move the products in ceramics" (CPO).*

After the I40 adoption, work systems include all I40 technologies which are integrated through the Internet of Things and produce a fully digital audit trail of operations keeping track of materials, workers, and machines who performed the actions. In this integrated assembly line, the intervention of the workforce is limited to the operation of a few machines whose tasks could not be so far automated. For the rest, the workforce supervises the automatic procedures and continuously improve the process by fine-tuning machines set-up to reduce defect rate and improve the quality of outputs. The workforce also provides feedback to technology experts to improve movements of mechanical arms. The I40 work systems allow increasing production quantity by 30% and the quality of products. Workers experienced a safer work and less muscular intensive tasks, which resulted in fewer work accidents and work illness.

*"Now we supervise the machines, we stay close to the machine, we check each operation, and we try to figure out how to improve the production" (Worker 2).*

*"Before we worked with a scorching temperature, now, these technologies improved our conditions mitigating the temperature. Now, I am supervising these machines producing ceramics. The work I do now is less manual and more mental. I can activate and deactivate the machine, and I try to figure out if there is a better way to let it work" (Worker 1).*

*"Now the role of workers is more mental, before was more manual. They use their knowledge to improve the process and digital competencies to control the technologies" (CPO).*

*"We increase the quality of products and improve the outcome quantity by 30%, while workers enjoy better work conditions, less accidents and illness" (CEO).*

## 5. Socio-Technical Interplay of Industry 4.0

Through the comparison of the work systems, the case shows the old and new interplay between workers and technology. In the traditional production process, the socio-technical interplay consisted in the realisation of production tasks where workers employ technologies. While in the I40 production line, the interplay concerns I40 technologies that perform production tasks entirely while the workforce performs supervision tasks and provides feedback for technology experts to improve technologies.

The socio-technical interplay in an I40 context realises when the workforce has distinctive competences to deal with I40 technologies and knowledge of crafting ceramics products. In our case, the experienced workforce already possesses knowledge in crafting ceramics products and acquires competences to work with technologies through vocational courses. While young workers have on-the-job training on the previous assembly line to gain knowledge in crafting ceramics and then they may pass to the I40 production line. Finally, technologies should be designed with the adaptive feature; that is, they afford to be reprogrammed in order to incorporate improvements. In our case, mechanical arms afford to change movements to increase the quality of products.

## 6. Discussion

Applying the socio-technical lenses on an I40 adoption, we noted how the role of workforce changes
in an I40 work system. Our study confirmed the literature that workers in an I40 context have a marginal role in the manual production activities [1, 11]. Still, their role passes from manual-skilled workers to a sort of maintenance worker of technology with advanced knowledge of the production process, which continuously optimises technology components. The resulting socio-technical interplay does not occur in production tasks but tasks related to the supervision of the I40 technologies. Therefore, the socio-technical lens shows how I40 adoption allows improving technical components improving the production process and how I40 adoption "humanises" social systems. The workforce can conduct activities more valuable based on knowledge and less muscular intensive in a safer workplace which result in the reduction of work accidents and work illness.

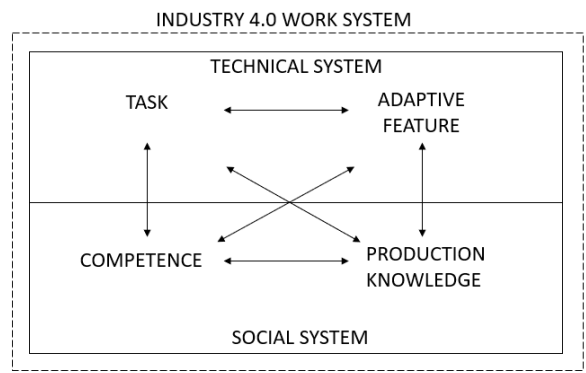

Figure 2 Socio-technical Framework for Industry 4.0

From the result of our study, we propose this socio-technical framework for I40 in Figure 2. This framework embraces four main components. The technical sub-system includes the tasks and the adaptive feature of technologies, while the social sub-system consists of the competence and production knowledge of the workforce. The conjoint optimisation of the work system occurs when workers possess proper competencies and production knowledge to manage I40 technologies. While the technical system supports to be reprogrammed to incorporate changes in their activities – we call this feature adaptive - which are proposed by workers that use their knowledge and competence to improve the production process. Therefore, we encourage researchers to validate our framework with quantitative or qualitative studies, particularly multiple case study or action research.

Moreover, each of these components deserves further investigation. Engineers should investigate the components of the technical system. For example, they examine how to automate tasks of the various production process and how to program technologies in order that they can incorporate changes in an effective way. While socio-technical researchers should study the proper competencies to use these technologies, and how organisations develop and manage the production knowledge of workers. The joint study of these two components also opens for a reflection related to the socio-technical tenet of the minimal critical specification, which refers to the management attitude to specify to the workers no more than was absolutely essential to operate in a work system [26, 27]. Thus, we propose the following question: "Is the minimal critical specification a mix of knowledge and competencies in an I40 context?" By answering this question, we can advance the knowledge of the nature of work in an I40 context.

Furthermore, the novel role of assembly line workers requires further studies. Literature shows that work practices that are changed by technologies may vary between small-medium and large organisations and the environmental context [28, 29]. Since we studied how the role of the workforce is changed in small-medium organisations in Italy, we encourage researchers to explore how changes the role of the workforce in large organisations and in organisations operating in a European context different from the Italian one, in America and Asia.

## 7. Conclusion

I40 initiative has a prominent technical perspective, and literature revealed a lack of a socio-technical perspective. To address this gap, our study presents an in-depth single case study of an Italian manufacturing organisation adopting I40 technologies, to which we apply the traditional socio-technical theory. We aim at investigating the socio-technical interplay of I40 work systems and the role of workers, which, according to the literature, is marginal down the assembly line. We find out that the socio-technical interplay occurs in tasks related to the supervision of the I40 technologies. The socio-technical interplay of I40 is based on the workforce competence to manage the technologies and the workforce knowledge of production. We conclude presenting a socio-technical framework of I40 and proposing future avenues for the socio-technical perspective of I40.